\documentclass[a4paper,12pt,final]{article}
\usepackage{amsmath,amsfonts,amssymb,amsthm,amstext,eucal,array}
\usepackage{graphicx,lscape}

\def\bea{\begin{eqnarray}} 
\def\eea{\end{eqnarray}}
\def\beann{\begin{eqnarray*}} 
\def\eeann{\end{eqnarray*}}
\def\be{\begin{equation}} 
\def\ee{\end{equation}}
\def\ba{\begin{array}} 
\def\ea{\end{array}}
\def\ben{\begin{enumerate}} 
\def\een{\end{enumerate}}

\def\4{\tilde }
\def\5{\bar }  
\def\6{\partial } 
\def\7{\hat } 

\def\G{\Gamma}

\def\cL{{\cal L}}
\def\cM{{\cal M}}
\def\cJ{{\cal J}}
\def\cN{{\cal N}}
\font\mybb=msbm10 at 10pt
\def\bb#1{\hbox{\mybb#1}}
\def\bR {\bb{R}}
\def\bZ {\bb{Z}}

\def\Ra {\text{R}_{\text{AdS}}}

\def\ha{\hat{\alpha}}
\def\hb{\hat{\beta}}
\def\hg{\hat{\gamma}}
\def\ua{\underline{\alpha}}
\def\ub{\underline{\beta}}
\def\umu{\underline{\mu}}
\def\unu{\underline{\nu}}
\def\hua{\hat{\underline{\alpha}}}
\def\hub{\hat{\underline{\beta}}}

\DeclareMathOperator{\AdS}{AdS}
\DeclareMathOperator{\dvol}{dvol}

\newcommand{\ISO}{\mathrm{ISO}}
\newcommand{\Og}{\mathrm{O}}
\newcommand{\SO}{\mathrm{SO}}
\newcommand{\SU}{\mathrm{SU}}

\begin{document}



\begin{titlepage}
\vfill
\begin{flushright}
WIS/38/02-AUG-DPP\\
\end{flushright}

\vfill

\begin{center}
\baselineskip=16pt
{\Large\bf Null orbifolds in AdS, Time Dependence and Holography}
\vskip 0.3cm
{\large {\sl }}
\vskip 10.mm
{\bf ~Joan Sim\'on }\\
\vskip 1cm
{\small
The Weizmann Institute of Science, Department of Particle Physics \\
Herzl Street 2, 76100 Rehovot, Israel \\
E-mail: jsimon@weizmann.ac.il}\\ 
\end{center}
\vfill
\par
\begin{abstract}
We study M/D-branes in a null-brane background. By taking a near horizon limit,
one is left with cosmological models in the corresponding Poincar\'e patches.
To deal with their usual horizons, we either extend these models to global
AdS or remain in the Poincar\'e patch and apply a T-duality transformation
whenever the effective radius of the compact dimension associated with the
null-brane probes distances smaller than the string scale. The first scenario
gives rise to null orbifolds in AdS spaces, which are described in detail.
Their conformal boundaries are singular. The second has a dual gauge theory 
description in terms of Super Yang-Mills in the null-brane background. The 
latter is a good candidate for a non-perturbative definition of string theory 
in a time-dependent background.
\end{abstract}
\vfill

\vfill

\end{titlepage}


\section{Introduction}

One of the most interesting questions in theoretical high energy physics
concerns the lessons that a quantum theory of gravity might teach us about 
the origin of the universe. It is thus natural to study time dependence
in the framework of string theory. Lately, there has been certain amount of
work in this direction. One can consider orbifolds and discrete quotients of 
Minkowski spacetime acting non-trivially on time \cite{horowitz,kt,seiberg,
paper1,vijay,costa,nekrasov,newcosta,joan1,LMS1,LMS2,fabinger}. Out of the
different possibilities, there is a single basic one preserving one half of
the spacetime supersymmetries, the null (parabolic) orbifold \cite{horowitz}
acting on $\bR^{1,2}$. This is a singular spacetime : it has fixed points, it
is non-Hausdorff and it has closed lightlike curves. These features are
resolved by the null-brane \cite{paper1,joan1,LMS2,fabinger} which is obtained 
by adding a shift, transverse to $\bR^{1,2}$, to the discrete quotient action.
The stability of these configurations \cite{albion,fabinger,Horpol} and the 
validity of perturbative techniques have also been partially discussed 
\cite{LMS1,LMS2,fabinger,Horpol}. Another possibility is that of coset models 
having a cosmological spacetime interpretation and allowing a conformal field 
theory description on the worldsheet \cite{hebrew,ben}. There are other 
approaches, such as considering certain double Wick rotations of previously 
known configurations \cite{AFHSWick, BRinWick, BRWick, CaiWick, GMWick, 
BLWWick}, S-branes \cite{GSSbranes, CGGSbranes,KMPSbranes, DKSbranes} and 
Sen's proposal \cite{SenRT,SenTE,Senlast} that dynamical rolling of the 
tachyon of open string field theory can lead to interesting cosmologies.

In this work, we shall be concerned with the first mentioned approach. The 
analysis made so far seems to indicate that the null-brane background is
stable to black hole formation when probed by particles for large shifts
and when the number of non-compact dimensions is big enough $(D>5)$
\cite{fabinger,Horpol}. However, our understanding of string dynamics in
these backgrounds is still poor, specially concerning the winding sector.
It is instructive to consider the open string sector by adding D-branes,
expecting to get some insight from their low energy gauge theory description
at weak coupling.

Our main goal will be to start exploring a possible non-perturbative
string cosmology scenario by analysing the null-brane background in the 
context of the AdS/CFT correspondence \cite{adscft,review}. It is thus 
necessary to know which M/D-branes exist in this vacuum, the amount of 
supersymmetry they preserve and their low energy closed string description
in terms of classical supergravity configurations. Most of these steps have 
been answered in ~\cite{paper2,paper3}, or consist of 
generalisations of the techniques developed there. We shall primarily be
concerned with branes in which the discrete quotient takes place along
their worldvolumes. Even though the dynamics of string theory in these
time dependent backgrounds is not understood, it is natural to study the
near horizon geometry of the corresponding M2/M5/D3 configurations. The 
null-brane carries an intrinsic scale, that we shall denote by $R$, which
gives the size of the neck of spacetime when moving from its contracting
phase to its expanding one. It is natural to keep it fixed when taking the
near horizon limit, which for D3-branes will be
\[
  \alpha^\prime\to 0 \quad , \quad \text{u}=\frac{\text{r}}{\alpha^\prime}
  \,, \text{g}_s\,,\text{R} \quad \text{fixed}\,.
\]
The resulting geometry gives rise to a cosmological model in the Poincar\'e
patch of $\AdS_5$. The latter has a similar geometrical interpretation
to the one in flat spacetime, but D3-branes, or more generally, any M/D-brane
that allows such a construction gives rise to horizons at u=0. There are two
main roads that one can follow to understand the meaning of these horizons :
\begin{itemize}
  \item[(i)] Extend the cosmological model by crossing the horizon. This leads 
  naturally to the analysis of orbifolds and discrete quotients of global
  AdS spaces.
  \item[(ii)] Since there is a compact dimension along the brane, whenever its
  effective radius probes distances smaller than the string scale, one is
  allowed to apply T-duality and use its T-dual description to understand the
  original horizon.
\end{itemize}

The first approach leads naturally to consider the analysis of null orbifolds
in AdS spaces \footnote{There have been previous works on orbifolds and 
discrete quotients of AdS spaces in the context of AdS/CFT, see
\cite{hormarolf,ads1,ads2,cai1}.}. This is because Lorentz 
transformations in the Poincar\'e patch are mapped into Lorentz 
transformations in $\bR^{2,p-1}$, the space in which we embed global $\AdS_p$.
On the other hand, translations in the first turned out to be null rotations 
(boost + rotation) in $\bR^{2,p-1}$, involving the ``second'' time.
To study the geometry, structure of singularities and conformal boundary of 
these discrete quotients, the embedding of $\AdS_p$ in $\bR^{2,p-1}$ turns out
to be very useful. As expected, the null orbifold of AdS spaces has fixed 
points and it is a non-Hausdorff space, as it happened in the flat spacetime 
orbifold. Somewhat more surprisingly, the analogue of the null-brane discrete
quotient in global $\AdS_p$ spaces $(p\geq 5)$ has no fixed points, but its 
conformal boundary is singular. It has fixed points, it is non-Hausdorff
and it has closed lightlike curves. It can be shown that this singular 
conformal boundary is equivalent to the conformal compactification of the 
null-brane manifold. These statements may lead us to conclude that if there is
any field theory dual to such cosmological scenarios in AdS, living on its
boundary, its base space must be a singular manifold.

The second approach is much more appealing. Borrowing some intuition from the
AdS/CFT correspondence, the cosmological model on the Poincar\'e patch of
$\AdS_5$ should be dual to a four dimensional Super Yang-Mills (SYM) gauge 
theory in the null-brane background. Thus, a priori, this scenario has some
perturbative description on the worldsheet, by extending the formalism
developed in \cite{LMS1} for the null orbifold to the open string sector
of the null-brane discrete quotient. It has a known low energy effective
supergravity description and a dual gauge theory. Furthermore, T-duality
can be studied both on the worldsheet and on supergravity. All in all, these 
might be the necessary ingredients to provide a non-perturbative definition
of string cosmology. We shall comment on the first steps to fulfill this
programme.

The organisation of the paper is as follows. In section 2, we explain the main
tools necessary to describe null orbifolds in AdS$_p$ spaces, $\forall$
$p\geq 3$. As an aside remark, we briefly comment on the embedding of the
BTZ black holes \cite{BTZ,henneaux} in higher dimensional AdS spaces.
In section 3, we embed the previous discussion in string/M-theory, by 
extending our analysis to include the q-forms relevant to 
$AdS_4\times S^7$, $AdS_7\times S^4$ and $AdS_5\times S^5$ spacetimes and 
by discussing supersymmetry considerations. In section 4, we 
comment on the corresponding orbifold construction in brane configurations
and their near horizon limits. In particular, we analyse the mapping of 
symmetries between the Poincar\'e patch and global AdS. In section 5, we 
discuss the analogue of the null-brane discrete quotient in AdS$_p$ spaces
$(p\geq 5)$ and discuss the singularities of their conformal boundary. 
In section 6, we comment on the gauge theory dual to the cosmological
models in the Poincar\'e patches, some T-duality properties of our
bulk configurations and on what we are still missing to fulfill this
programme. Some explicit computations to prove the amount of supersymmetry 
preserved by the discrete quotients discussed in the body of the paper are 
performed in an appendix.

\textsc{Note added.}  While working on this project, we received \cite{akisav}
which discusses similar cosmological scenarios and their holographic duals
in terms of time-dependent non-commutative field theories. Notice that in
their scenario, the shift that gives rise to the null-brane is transverse
to the brane (D2-brane in the forementioned reference), and so it describes
a physically inequivalent model to the ones considered in this work.

\section{Null orbifolds in AdS spaces}

The p-dimensional anti-de Sitter space $(AdS_p)$ is a maximally symmetric
space of constant negative curvature. It can be represented as the
hyperboloid
\begin{equation}
  \begin{aligned}
    \text{AdS}_p & \hookrightarrow \bR^{2,p-1} \\  
    -u^2 - v^2 + & \sum_{i=1}^{p-1} (x^i)^2 = -(\Ra)^2 \quad ,
  \end{aligned} 
 \label{embedding}
\end{equation}
in the flat (p+1)-dimensional space $\bR^{2,p-1}$. By construction, the space
has the isometry group $\Og(2,p-1)$ \footnote{When embedding these spaces
in string theory, the presence of fluxes will restrict this isometry group
to $\SO(2,p-1)$.}. 

Equation \eqref{embedding} can be solved by setting
\begin{equation}
  \begin{aligned}
    u&=\Ra\,\cosh\rho\,\cos\tau \quad , \quad 
    v=\Ra\,\cosh\rho\,\sin\tau \\
    x^i &=\Ra\,\sinh\rho\,\Omega^i \quad (i=1,\dots , p-1;\quad \sum_i
    (\Omega^i)^2=1)\,,
  \end{aligned}
 \label{global}
\end{equation}
where $\{\Omega^i\}$ parametrise a unit (p-2)-sphere.
The induced metric on the hyperboloid turns out to be
\begin{equation}
  g_{AdS_{p}} = (\Ra)^2\left[-(\cosh\rho)^2 (d\tau)^2 + (d\rho)^2 +
  (\sinh\rho)^2 g_{S^{p-2}}\right]
 \label{globalmetric}
\end{equation}
where $g_{S^{p-2}}$ stands for the metric on a unit $\text{S}^{p-2}$ sphere.
As usual in the physics literature, we shall refer to $\text{AdS}_p$ to the
universal covering space of the above hyperboloid \eqref{embedding} in which
the global timelike coordinate $\tau$ has been unwrapped (i.e. take
$-\infty < \tau < \infty$).

Any Killing vector $\xi$ defines a one parameter subgroup of isometries
whose action in $\text{AdS}_p$ is given by
\begin{equation}
  \text{P} \,\rightarrow\, e^{t\xi}\text{P} \quad \forall\,\text{P}\in
  \text{AdS}_p
 \label{u1action}
\end{equation}
In this section, we shall be concerned with the subgroup generated by
\footnote{The existence of such a generator requires us to work with
$\text{AdS}_p$ $p\geq 3$.}
\begin{equation}
  \xi^\pm =  \frac{1}{\sqrt{2}}\left[u\partial_x + x\partial_u
  \pm \left(y\partial_x - x\partial_y\right)\right] 
  = x^\pm\partial_x + x\partial_\mp ~,
 \label{nullkilling}
\end{equation}
where $x^\pm = (u\pm y)/\sqrt{2}$. Notice that $\xi^\pm$ is the linear
combination of a boost in the ux-plane and a rotation in the yx-plane.
The quotient manifold $\text{AdS}_p/\Gamma^\pm $ is obtained by identifying
points under the action \eqref{u1action} whenever $t$ is an integer multiple
of a basic discrete step. 

In the following, we shall discuss the geometry of these quotient
manifolds both for global $AdS_p$, and for particular patches of it,
the second option giving rise to the embedding of the BTZ black hole
~\cite{BTZ,henneaux} in higher dimensional $AdS_p$ spaces $(p\geq 4)$.

\subsection{Global AdS}

The embedding $AdS_p\hookrightarrow \bR^{2,p-1}$ provides us with the most 
natural way to study the null orbifold $AdS_p/\Gamma^+$. Introduce coordinates
$\{x^+\,,x^-\,,x\}$ on $\bR^{1,2}\subset \bR^{2,p-1}$ $(p\geq 3)$,
being defined as above and assembled them into a column vector $X$. 
The generator $g_0$ of the orbifold acts as
\begin{equation}
  X = 
  \begin{pmatrix}
    x^+ \\ x \\ x^-
  \end{pmatrix}
  \,\rightarrow \, g_o\cdot X = e^{t\cJ}\,X =
  \begin{pmatrix}
    x^+ \\ x + tx^+ \\ x^- + tx + \frac{1}{2}t^2 x^+
  \end{pmatrix}
  ~;
  \cJ =
  \begin{pmatrix}
    0 & 0 & 0 \\
    1 & 0 & 0 \\
    0 & 1 & 0
  \end{pmatrix}
\end{equation}

The above definition coincides with $\bR^{1,2}/\Gamma^+$, but it certainly
applies to $AdS_p/\Gamma^+$ since the action of $\Gamma^+$ leaves 
\eqref{embedding} invariant. Following the discussion in 
\cite{joan1,LMS1}, the action of $\Gamma^+$ on the embedding space
$\bR^{2,p-1}$ has a subset of fixed points located at $x^+=x=0$. Furthermore,
$\bR^{2,p-1}/\Gamma^+$ is a non-Hausdorff space, as can be seen by inspection
of the $x^+=0$ subspace. To learn about $AdS_p/\Gamma^+$, one evaluates
\eqref{embedding} in these singular subspaces. Denoting $x^{p-2}=x$ and
$x^{p-1}=y$, the set of fixed points in $AdS_p/\Gamma^+$ is described by
\[
  v_{\pm} = \pm \sqrt{(\Ra)^2 + \sum_{i=1}^{p-3} (x^i)^2}~,
\]
consisting of two disconnected branches, whereas the loci where $\|\xi^+\|^2
=(x^+)^2=0$ is given by
\[
   v_{\pm} = \pm \sqrt{(\Ra)^2 + \sum_{i=1}^{p-3} (x^i)^2 + (x)^2}~.
\]
The fact that $\{x^+=0\}\in\text{AdS}_p$ $\forall\,\text{x}$ allows us to 
conclude that $AdS_p/\Gamma^+$ is also a non-Hausdorff space.

Working in global coordinates \eqref{global} and picking $\Omega^x=\cos\theta$
and $\Omega^y=\sin\theta\cos\psi\,$ $\,0\leq \theta\,,\psi < \pi$ , for 
concreteness, we can describe the set of fixed points by
\begin{equation}
  \begin{aligned}
    \rho=0 & \quad , \quad \tau=\frac{\pi}{2} + \pi\,\text{n} \quad \text{n}\in
    \bZ \quad \forall\,\vec{\Omega}\in \text{S}^{p-2} \\
    \theta=\frac{\pi}{2} & \quad , \quad \cos\tau=-\tanh\rho\cos\psi \quad
    \forall\,\vec{\Omega}\in \text{S}^{p-4}_{\sin\psi}
  \end{aligned}
 \label{nullfix}
\end{equation}
where $\text{S}^{p-4}_{\sin\psi}$ stands for a (p-4)-sphere of radius
$\sin\psi$. 

It is particularly interesting to study the action of $\Gamma^+$ on the
conformal boundary of $\text{AdS}_p$. It is well-known that $\text{AdS}_p$
can be conformally mapped to one half of the Einstein static universe.
This can be seen by introducing a new coordinate $\varphi$ related to $\rho$
by
\[
  \tan\varphi = \sinh\rho \quad (0\leq\varphi< \frac{\pi}{2})~.
\]
The $\text{AdS}_p$ metric is conformal to
\begin{equation}
  g_{E(p)} = -(d\tau)^2 + (d\varphi)^2 + (\sin\varphi)^2g_{S^{p-2}}~.
\end{equation}
It is now manifest that the spacelike hypersurfaces of constant $\tau$ are
(p-1)-hemispheres, whose equator $(\varphi=\pi/2)$ is a boundary with
topology $\text{S}^{p-2}$. The boundary of the full $\text{AdS}_p$ spacetime
is thus $\bR\times \text{S}^{p-2}$, and located at $\varphi=\pi/2$
$(\rho\to\infty)$. 

Moving to $\AdS_p/\Gamma^+$, one can derive the set of fixed points
on the boundary by evaluating \eqref{nullfix} at $\rho\to\infty$, giving rise
to :
\[
  \theta=\frac{\pi}{2} \quad , \quad \tau= \pi \pm \psi \,\,(\text{mod}\,2\pi)
  \quad \forall\,\vec{\Omega}\in \text{S}^{p-4}_{\sin\psi}
\]
Using similar arguments as before, one concludes that the boundary is
non-Hausdorff due to the specific features of the hypersurface
\[
  \cos\tau=-\sin\theta\cos\psi ~.
\]

The above analysis suggests that if there is any field theory dual living
on the boundary, it must be defined on a singular base space $\cM$, which 
we shall later identify with the conformal compactification of the null-brane
manifold. Before discussing this point, we shall first find a useful local
description for the orbifold $\text{AdS}_p/\Gamma^+$. Inspired by
the embedding $AdS_p\hookrightarrow \bR^{2,p-1}$, it is natural to
introduce a local adapted coordinate system $\{z^+\,,z\,,z^-\}$
\begin{equation}
  \begin{aligned}
    x^+ &= z^+ \\
    x &= z\,z^+ \\
    x^- &= z^- + \frac{1}{2}z^+\,z^2
  \end{aligned}
 \label{adapted}
\end{equation}
in which $\xi^+=\partial_z$, so that the identifications become shifts in the 
$z$ variable
\begin{equation}
  (z^+\,,z\,,z^-)\sim (z^+\,,z+2\pi\,,z^-)
\end{equation}
in some convenient normalization. It is important to keep in mind that
the coordinate system $\{z^+\,,z\,,z^-\}$ defined in \eqref{adapted} breaks
down at $z^+=0$. Rewriting \eqref{embedding} in terms of
$\{z^+\,,z\,,z^-\}$
\begin{equation}
  -2z^+z^- -v^2 + \sum_{i=1}^{p-3} (x^i)^2 = -(\Ra)^2 ~,
 \label{adaptemb}
\end{equation}
one discovers that it is independent of the compact variable $z$. One thus 
learns that for $z^+\neq 0$, the metric for the null orbifold 
$\text{AdS}_p/\Gamma^+$ is that of
\begin{equation}
  g_{AdS_p/\Gamma^+} = g_{AdS_{p-1}} + (z^+)^2 (dz)^2 ~,
 \label{nullads}
\end{equation}
that is, some $S^1$ fibration over $AdS_{p-1}$. Notice that $\text{AdS}_{p-1}$
has the same radius $\Ra$ as the original $\text{AdS}_p$ and that the effective
radius $(z^+)$ of the compact direction (z) belongs to $\text{AdS}_{p-1}$,
for $z^+\neq 0$. The time dependence of the solution becomes manifest
when one expresses $z^+$ in terms of the global parametrisation \eqref{global}
of $\AdS_{p-1}$.

Following the discussion on the conformal boundary for $\text{AdS}_p$, it
is straightforward to determine the conformal boundary for 
$\text{AdS}_p/\Gamma^+$. Using the global parametrization for
$\text{AdS}_{p-1}$ appearing in \eqref{nullads}, this metric is conformal
to
\begin{equation}
  g = g_{E(p-1)} + \frac{1}{2}\left(\cos\tau+\sin\tilde{\varphi}\,\Omega^y
  \right)^2 (dz)^2 ~,
 \label{boundary}
\end{equation}
where the non-compact coordinate $\rho$ in $\text{AdS}_{p-1}$ was related to
$\tilde{\varphi}$ through $\tan\tilde{\varphi}=\sinh\rho$. Once more, the
conformal boundary is at $\tilde{\varphi}=\pi/2$, since $S^1$ does not have  
one. Notice that $\Omega^y$ is just indicating the direction in which
the null rotation identification was done. Thus the metric on the conformal 
boundary is given by
\begin{equation}
  g_{\partial\text{AdS}_p/\Gamma^+} = -(d\tau)^2 + g_{S^{p-3}} +
  \frac{1}{2}\left(\cos\tau+\Omega^y\right)^2 (dz)^2 ~.
 \label{confbound}
\end{equation}

We would finally like to prove that the conformal boundary of 
$\text{AdS}_p/\Gamma^+$ is equivalent to the conformal compactification
of $\bR^{1,p-2}/\Gamma^+$. First of all, we introduce adapted coordinates
for $\bR^{1,2}\subset \bR^{1,p-2}$ where $\Gamma^+$ acts non-trivially, just
as in \eqref{adapted}. By standard manipulations for studying the conformal
infinity in flat Minkowski spacetime, the metric on $\bR^{1,p-2}/\Gamma^+$
is conformal to
\begin{equation}
  g = -(d\tau^\prime)^2 + (d\theta)^2 + (\sin\theta)^2g_{S^{p-4}} +
  \frac{1}{2}\left(\sin\tau^\prime + \sin\theta\, \Omega^{\tilde{y}}\right)^2
  (dz)^2~.
 \label{confcomp}
\end{equation}
Of course, the above description also breaks down when $\tilde{y}^+=0$.
It is though manifest that both \eqref{confbound} and \eqref{confcomp}
are equivalent if $\Omega^y = \sin\theta\,\Omega^{\tilde{y}}$ and
$\tau^\prime=\tau + \pi/2$, that is, if the directions of the corresponding
null rotations are conveniently identified. Strictly speaking, the above 
explicit computation shows that the conformal boundary for 
$\text{AdS}_p/\Gamma^+$ in the adapted coordinate system 
$\{z^\pm\,,z\}$ is equal to the conformal compactification of 
$\bR^{1,p-2}/\Gamma^+$ in the adapted coordinates $\{\tilde{y}^\pm\,,z\}$. 
On the other hand, even though both sets of coordinates break down in 
different sets of points, both sets actually coincide when restricting them 
to the confomal boundary and compactification, respectively. This allows us 
to state that the conformal boundary of $\text{AdS}_p/\Gamma^+$ is equal to 
the conformal compactification of $\bR^{1,p-2}/\Gamma^+$.

Notice that concentrating on the hypersurface defined by the fixed values
$\rho=\rho_0$, $\theta=\theta_0$ and zooming the region close to the 
singularity $\tau_0$, by expanding $\tau= \tau_0 + \delta\tau$, 
$\delta\tau\ll1$, one recovers the double cone scenario advocated in 
\cite{LMS1}
\begin{equation}
  g_{AdS_p/\Gamma}\sim \frac{R^2}{(\cos\tilde\theta)^2}\left\{
  -(d\delta\tau)^2 + (\sin\tilde\theta_0)^2(\sin\theta_0)^2\,g_{S^{p-4}}
  + \frac{1}{2}(\sin\tau_0)^2(\delta\tau)^2 (d\tilde{y})^2\right\}~.
\end{equation}

\subsection{BTZ black holes}

The extremal 2+1 BTZ black hole \cite{BTZ} can be understood as the quotient
of a subset of $AdS_3$ by the discrete action generated by 
$\xi^+$. See \cite{henneaux} for a more detailed discussion.
What we would like to address here concerns the embedding of such a 
construction in higher dimensional $AdS_p$ $(p\geq 4)$ spaces. To do so, we
set
\[
  -u^2 - v^2 + x^2 + y^2 = -\rho^2 ~,
\]
and inserting it back into \eqref{embedding}, we learn $\rho\in[R\,,\infty)$.
It is thus natural to introduce global coordinates
\begin{equation}
  \begin{aligned}
    \rho &= R\cosh\mu \\
    x^i &= R\sinh\mu\,\hat{x}^i
  \end{aligned}
 \label{mudef}
\end{equation}
where $\hat{x}^i$ parametrise a (p-4)-sphere. Notice that the range of the 
coordinate $\mu$ introduced in \eqref{mudef}, depends on the dimension
of the full $AdS_p$ space under consideration. In particular,
\begin{equation}
  \begin{aligned}
    p=4 & \quad \Rightarrow \quad -\infty < \mu < \infty \\
    p>4 & \quad \Rightarrow \quad 0 \leq \mu < \infty ~.
  \end{aligned}
\end{equation}
On the other hand, we shall parametrise the $\{u\,,v\,,x\,,y\}$ in terms
of Poincar\'e coordinates, which now will depend on the point $\mu$ 
\begin{equation}
  y+u = \frac{R\cosh\mu}{z} \quad ,\quad x =\frac{R\cosh\mu}{z}\,\beta 
  \quad v = -\frac{R\cosh\mu}{z}\,\gamma ~.
 \label{poincare}
\end{equation}

It is precisely when one restricts to a single patch of $AdS_3$ with a
definite sign of the z coordinate ($z>0$, for instance), that by modding
out the corresponding spacetime by the discrete action generated by
$\xi^+$ one expects to embed the BTZ black hole in $AdS_p$.
This is explicitly checked by rewriting the induced metric on 
\eqref{embedding} in the coordinate system defined in \eqref{mudef}
and \eqref{poincare}. The result is given by
\begin{equation}
  g_{AdS_p/\Gamma} = R^2\left\{(\cosh\mu)^2 g_{\text{BTZ}} + (d\mu)^2 +
  (\sinh\mu)^2 g_{S^{p-4}}\right\}~, 
 \label{btzemb}
\end{equation}
where $g_{\text{BTZ}}$ stands for the 2+1 massless extremal BTZ black hole 
metric,
\begin{equation}
  g_{\text{BTZ}} = -r^2 (dt)^2 + \frac{(dr)^2}{r^2} + r^2 (d\phi)^2
 \label{btz}
\end{equation}
which was written in the standard coordinates $t=\gamma$, $\phi=\beta$
and $r=z^{-1}$, all of them being dimensionless. On the other hand,
$g_{S^{p-4}}$ stands for the metric on a unit sphere of dimension $p-4$.

It is worthwhile pointing out that the above embedding of the extremal 
BTZ black hole in $AdS_p$ $(p\geq 4)$, can be straightforwardly generalised 
to the non--extremal ones. If instead of considering the action generated by 
$\xi^+$, one would have considered the action generated by any of the Killing 
vectors spanning $\SO(2,2)$ giving rise to non-extremal BTZ black holes
(see \cite{henneaux}), then one would have parametrised $\{u^i\}=
\{u\,,v\,,x\,,y\}$ by
\[
  u^i = R\cosh\mu\,\hat{u}^i \quad , \quad -\hat{u}^2-\hat{v}^2
  + \hat{x}^2 + \hat{y}^2 = -1 ~.
\]
The choice of $\{\hat{u}^i\}$ is directly related with the choice of
discrete quotient (or Killing vector). The discussion on the remaining 
coordinates $\{x^i\}$ is as before. All in all, one derives the metric 
involving a non-extremal BTZ black hole $(\tilde{g}_{BTZ})$ embedded in 
$AdS_p$ for $p\geq 4$ :
\begin{equation}
  g_{AdS_p/\Gamma} =R^2\left\{(\cosh\mu)^2 \tilde{g}_{\text{BTZ}} + (d\mu)^2 +
  (\sinh\mu)^2 g_{S^{p-4}}\right\} ~.
\end{equation}

\section{Embedding in String/M-theory}

Even though there are many known examples for embeddings of AdS spaces of
different dimensions in string theory, we shall concentrate on the maximally
supersymmetric ones. These are of the form 
$\text{AdS}_{p+2}\times\text{S}^{D-p-2}$ and can be thought of near horizon
geometries of the M2-, M5- and D3-brane configurations. The values of $p$ and
$D$ are listed in Table \ref{adsxs} along with the ratio 
$\iota=\text{R}_{\text{AdS}_{p+2}}/\text{R}_{\text{S}^{D-p-2}}$ of the radii
of curvature of the two factors.

\begin{table}[!ht]
  \begin{center}
    \setlength{\extrarowheight}{5pt}
    \begin{tabular}{|c|c|c|c|}
      \hline
      Brane & p & D & $\iota$ \\
      \hline
      M2 & 2 & 11 & $\frac{1}{2}$ \\
      D3 & 3 & 10 & 1 \\
      M5 & 5 & 11 & 2 \\[3pt]
      \hline
    \end{tabular}
    \vspace{8pt}
    \caption{Dimensions and radii of curvature}
    \label{adsxs}
  \end{center}
\end{table}

The metric on the direct product $\text{AdS}_{p+2}\times\text{S}^{D-p-2}$ is
\[
  g = g_{\text{AdS}_{p+2}} + (\text{R}_{\text{S}})^2\,g_{\text{S}^{D-p-2}}~.
\]
These configurations do also carry fluxes. Using the same conventions as in
~\cite{ppwave}, these are summarised below :
\begin{equation}
  \begin{aligned}
    \text{AdS}_4\times\text{S}^7\quad : F_4 &= \frac{3}{\Ra}\dvol\left(
    \text{AdS}_4\right) \\
    \text{AdS}_5\times\text{S}^5\quad : F_5 &= \frac{1}{2\Ra}\left(
    \dvol\left(\text{AdS}_5\right) + \dvol\left(\text{S}^5\right)\right) \\
    \text{AdS}_7\times\text{S}^4\quad : F_4 &= \frac{6}{\Ra}\dvol\left(
    \text{S}^4\right) ~.
  \end{aligned}
 \label{fluxes}
\end{equation}

It is straightforward to extend our previous analysis to these supersymmetric
vacua. By construction, the action of $\Gamma^+$ is restricted to the
$\text{AdS}_{p+2}$ part of the direct product, giving rise to
$\left(\text{AdS}_{p+2}/\Gamma^+\right)\times\text{S}^{D-p-2}$. Therefore,
in the local adapted coordinate system \eqref{adapted}
\begin{equation}
  g_{\left(\text{AdS}_{p+2}/\Gamma^+\right)\times\text{S}^{D-p-2}} = 
  g_{\text{AdS}_{p+1}} + (z^+)^2 (dz)^2 + 
  (\text{R}_{\text{S}})^2\,g_{\text{S}^{D-p-2}}~.
 \label{f1}
\end{equation}

Concerning the field strengths \eqref{fluxes}, due to its geometrical content,
it is easy to work out their local expressions as
\begin{equation}
  \begin{aligned}
    \left(\text{AdS}_4\right)/\G^+\times\text{S}^7\quad : F_4 &= 
    \frac{3}{\Ra}z^+\,\dvol\left(
    \text{AdS}_3\right)\wedge dz \\
    \left(\text{AdS}_5\right)/\G^+\times\text{S}^5\quad : F_5 &= 
    \frac{1}{2\Ra} z^+\,\left(
    \dvol\left(\text{AdS}_4\right)\wedge dz + \dvol\left(\text{S}^5\right)
    \right) ~,
  \end{aligned}
 \label{nullfluxes}
\end{equation}
whereas $F_4$ is left unmodified for $\text{AdS}_7\times\text{S}^4$. Remember
that these expressions break down at $z^+=0$ and notice that their fluxes
over $\text{S}^{D-p-2}$ are unchanged.

Whenever one considers an orbifold of any supersymmetric vacua, there is an
issue concerning the amount of supersymmetry preserved. Just as there is
a known correspondence among parallel spinors in $\bR^{n+1}$ and Killing
spinors in the n-sphere $\left(\text{S}^n\hookrightarrow\bR^{n+1}\right)$
embedded in it ~\cite{jose1}, one can prove that there is an analogous
correspondence between parallel spinors in $\bR^{2,p-1}$ and Killing spinors
on $\text{AdS}_p\hookrightarrow\bR^{2,p-1}$ \footnote{This correspondence
was found in collaboration with J.M. Figueroa-O'Farrill and will be discussed
in \cite{paper5}.}. In this way, since the null orbifold breaks half of the
supersymmetries in $\bR^{2,p-1}$, we learn the corresponding orbifolds
of the maximally supersymmetric AdS spaces preserve one half of the original
supersymmetries. In any case, we give an explicit proof of this fact in the
appendix.

\section{Null orbifolds in branes}

When dealing with null orbifolds in flat spacetime, their singularities
are naturally smoothed by modifying the action by which one quotients
the original manifold. In other words, one considers a different Killing vector
from the one given in \eqref{nullkilling}. This gives rise to the null-brane
discrete quotient \cite{paper1,joan1} which is associated with the Killing 
vector
\[
  \xi = R\partial_z + \xi^\pm ~,
\]
$z$ being some spacelike direction orthogonal to the three dimensional
lorentzian space where $\xi^\pm$ acts.

As emphasized in ~\cite{joan1} and extensively discussed in \cite{paper2},
a sufficient condition for a brane-like configuration to exist in such
string vacuum (null-brane) is to have an $\ISO(1,3)$ isometry
subgroup. This is the case in which we are going to concentrate, and so it
will not apply for M2-branes. The structure of singularities is analogous
to the one in flat spacetime, since the branes under consideration are flat.
The main difference though, is the existence of horizons associated
with brane throats.

We shall concentrate on D3-branes, even though a similar discussion would
apply for M5-branes. Following \cite{paper1}, the classical supergravity
configuration describing D3-branes in a null-brane vacuum has a ten dimensional
metric
\begin{multline}
    g = V^{-1/2}\left\{-2dy^+\,dy^- + dy^2 + \left(1+
    \left(\frac{y^+}{R}\right)^2\right)\,dz^2 \right. \\
    \left. + 2\,dz\left(\frac{y^+}{R}dy -\frac{y}{R}dy^+\right)\right\}
    + V^{1/2}\left(dr^2 + r^2\,g_{S^5}\right)~,
 \label{d3null}
\end{multline}
with constant dilaton and the corresponding self-dual five form flux.
The function $V=V(r)$ is the usual harmonic function
\[
  V(r)= 1 + \frac{Q^4}{r^4} \quad , \quad Q^4 = 2\pi^{3/2}\,
  \text{g}^2_{\text{YM}}\text{N}\,(\alpha^\prime)^2~.
\]
Taking the near horizon limit
\[
  \alpha^\prime\to 0 \quad , \quad \text{u}=\frac{\text{r}}{\alpha^\prime}
  \,, \text{g}_s\,,\text{R} \quad \text{fixed}\,,
\]  
one is left with the cosmological model
\begin{multline}
  g=\alpha^\prime\left\{\frac{\text{u}^2}{\sqrt{\lambda}}
  \left(-2dy^+\,dy^- + dy^2 + \left(1+\left(\frac{y^+}{R}\right)^2\right)
  \,dz^2 \right.\right. \\
  \left. \left. + 2\,dz\left(\frac{y^+}{R}dy -\frac{y}{R}dy^+\right)\right) 
  + \frac{\sqrt{\lambda}}{\text{u}^2} d\text{u}^2 +
  \sqrt{\lambda}\,g_{S^5}\right\}~,
 \label{cosmohor}
\end{multline}
where $\lambda=2\pi^{3/2}\,\text{g}^2_{\text{YM}}\text{N}$ and the 
corresponding self-dual five form fluxes. The above spacetime just covers
the Poincar\'e patch $(0<\text{u}<\infty)$. It is manifest that the effective
radius of the compact dimension $z$ vanishes on the horizon $(\text{u}=0)$.
At this stage, one may decide to maximally extend the above discrete
quotient to global AdS, by crossing the horizon, or whenever such an
effective radius becomes smaller than the string scale, one may
apply T-duality. This second possibility will be postponed to the last section.

In order to carry on the first possibility, we need to identify how the 
generator of the discrete quotient in the Poincar\'e patch $(\xi=R\partial_z
+ \xi^\pm)$ acts in global AdS. We shall do this for $\AdS_{p+2}$. Let us
remind the relation among the coordinates parametrising the Poincar\'e patch
$(y^\mu\,,u)$ and the ones describing the embedding
$\text{AdS}_{p+2}\hookrightarrow \bR^{2,p+1}$ \eqref{embedding} :
\begin{equation}
  \begin{aligned}
    X^\mu &= \frac{u}{R_{\text{AdS}}}x^\mu \quad (\mu=0,1,\dots ,p-1) \\
    X^{p+1} &= \frac{R^2_{\text{AdS}}}{2u}\left[1+\frac{u^2}{R^2_{\text{AdS}}}
    \left(1+\frac{\eta_{\mu\nu}y^\mu y^\nu}{R^2_{\text{AdS}}}\right)\right] \\
    X^p &= \frac{R^2_{\text{AdS}}}{2u}\left[1-\frac{u^2}{R^2_{\text{AdS}}}
    \left(1-\frac{\eta_{\mu\nu}y^\mu y^\nu}{R^2_{\text{AdS}}}
    \right)\right]~,
  \end{aligned}
 \label{patchmap}
\end{equation}
where we identified $X^{p+1}=u$ and $X^0=v$, without losing generality.

Using the identities :
\[
  \frac{\partial X^\mu}{\partial y^\nu} = \frac{u}{\Ra}\delta^\mu_\nu
  \quad , \quad \frac{\partial X^{p+1}}{\partial y^\nu} = 
  \frac{\partial X^p}{\partial y^\nu}=\frac{\eta_{\nu\mu}X^\mu}{\Ra}
  \quad , \quad X^{p+1}-X^p=u ~,
\]
it is manifest that
\begin{equation}
  \begin{aligned}
    P_\mu &= \partial_\mu \,\rightarrow\, \frac{\sqrt{2}}{\Ra}
    \left(X_\mu\partial_+ - X_+\partial_\mu\right) \\
    L_{\mu\nu} &= y_\mu\partial_\nu +  y_\nu\partial_\mu
     \,\rightarrow\, X_\mu\partial_\nu +  X_\nu\partial_\mu
  \end{aligned}
 \label{match}
\end{equation}
where $X^\pm=\left(X^{p+1}\pm X^p\right)/\sqrt{2}$.

The above computation tells us that Lorentz transformations belonging to
$\SO(1,p)$ in the Poincar\'e patch are mapped into Lorentz transformations in 
$\SO(2,p+1)$ in global AdS, which in particular justifies our previous
analysis on null orbifolds in AdS spaces. Furthermore, translations
$(\Ra\partial_\mu)$ in the Poincar\'e patch are mapped to null rotations
in $\SO(2,p+1)$ involving ``the second time'' in $\bR^{2,p+1}$. To be more
precise,
\[
  \Ra\partial_\mu \,\rightarrow\, \sqrt{2}\xi^-
\]
as defined in \eqref{nullkilling}. This analysis strongly suggests to
analyse the geometrical structure of $\text{AdS}_p$ $(p\geq 5)$ modded out
by the discrete action generated by $\xi=\xi^+(u,x,y) + \xi^-(v,s,r)$.
This is the subject of the next section.

\section{Null-brane discrete quotients in AdS spaces}

Let us parametrise $\text{AdS}_p$ $(p\geq 5)$ as the hyperboloid
\begin{equation}
  -u^2 + x^2 + y^2 -v^2 + r^2 + s^2 + \sum_i^{p-5}(x^i)^2 = -(\Ra)^2 ~,
 \label{newembed}
\end{equation}
and study the quotient manifold $\text{AdS}_p/\Gamma^{+-}$ obtained by
a discrete identification of points using the action $\Gamma^{+-}$
associated with the Killing vector
\begin{equation}
  \xi = \xi^+(u\,,x\,,y) + \xi^-(v\,,s\,,r)
 \label{nullkilling2}
\end{equation}
Introducing coordinates $\{x^\pm\,,x\}$ and $\{s^\pm\,,s\}$ to describe both
$\bR^{1,2}$ factors where $\Gamma^{+-}$ acts non-trivially and assembling
them into a column vector $X$, the generator $g_0$ of the orbifold acts as
\begin{equation}
  X = 
  \begin{pmatrix}
    x^+ \\ x \\ x^- \\  s^- \\ s \\ s^+
  \end{pmatrix}
  \quad \rightarrow \quad g_o\cdot X = e^{t\cJ}\,X =
  \begin{pmatrix}
    x^+ \\ x + tx^+ \\ x^- + tx + \frac{1}{2}t^2 x^+ \\
    s^- \\ s + ts^- \\ s^+ + ts + \frac{1}{2}t^2 s^-
  \end{pmatrix}
\end{equation}
where the matrix $\cJ$ is given by
\begin{equation*}
  \cJ =
  \begin{pmatrix}
    0 & 0 & 0 & 0 & 0 & 0 \\
    1 & 0 & 0 & 0 & 0 & 0  \\
    0 & 1 & 0 & 0 & 0 & 0 \\
    0 & 0 & 0 & 0 & 0 & 0 \\
    0 & 0 & 0 & 1 & 0 & 0 \\
    0 & 0 & 0 & 0 & 1 & 0
  \end{pmatrix}~.
\end{equation*}

It is clear that such an action has fixed points in the embedding space
$\bR^{2,p-1}$. These are located at $x^+=x=s^-=s=0$. The question is whether 
these points belong to our original manifold, $\text{AdS}_p$.
It is manifest that they do not satisfy \eqref{newembed} for any finite
$\{x^-\,,s^+\}$ value, thus allowing us to conclude the absence of fixed 
points under the action of $\Gamma^{+-}$ in $\text{AdS}_p$. Analogously, the 
set of points in $\bR^{2,p-1}$ where $\|\xi\|^2=0$, that is, $s^-=x^+=0$, 
does not belong to $\text{AdS}_p$. One thus concludes that the vector field 
$\xi$ \eqref{nullkilling2} is spacelike everywhere in $\text{AdS}_p$.
The corresponding quotient manifold $\text{AdS}_p/\Gamma^{+-}$ must be smooth.

Whenever one constructs quotient manifolds, the non-existence of closed
causal curves (CCC's) in it is non-trivial. Notice that the distance between
a point and its n'th image in our construction is $|t\,n|\sqrt{(x^+)^2 +
(s^-)^2}$. Using the same argument as before, such distance is always 
positive in $\text{AdS}_p$. Since the integral curves of the Killing vector 
\eqref{nullkilling2} are geodesic, the above distance is geodesic. Therefore,
$\text{AdS}_p/\G^{+-}$ has no CCC's.

Working in global coordinates \eqref{global} allows us to study the action
of $\Gamma^{+-}$ on the conformal boundary of $\text{AdS}_p$. It is then easy 
to check that indeed there are no fixed points for finite $\rho$ (the 
non-compact spacelike direction in global AdS), but that there is a curve of
fixed points on an infinite cylinder of maximal circle base extending along
global time $\tau$ in AdS and lying in the ry-plane. It is also on this curve
that $\|\xi\|^2$ vanishes. To understand the origin of these singularities,
consider the conditions $x^+=s^-=0$ in global coordinates plus the fundamental
trigonometric identity. This gives the constraint
\[
  \left(\tanh\rho\right)^2\left[(\Omega^r)^2 + (\Omega^y)^2\right] = 1 ~,
\]
which can only be satisfied for $\rho\to\infty$ (conformal boundary)
and $(\Omega^r)^2 + (\Omega^y)^2 = 1$. The latter defines a maximal circle
in the ry-plane and enforces $\Omega^i=0$ $\forall$ i$\neq$ r,s. Describing
this circle by $\varphi$ $(0\leq\varphi<2\pi)$, the curve on the cylinder is
given by
\[
  \tau = -\varphi \quad (\text{mod}\,\pi)~.
\]
This discussion allows us to state that if there is any field theory dual
description, it would be defined in a singular base space.

As in the discussion of null orbifolds in previous sections, it is useful
to introduce a local adapted coordinate system 
$\{z^\pm\,,\omega^\pm\,,z\,,\phi\}$ to describe the geometry of
$\text{AdS}_p/\Gamma^{+-}$
\begin{equation}
  \begin{aligned}
    s^- = \omega^- \quad , & \quad x^+ = z^+ \\
    s = \phi\,\omega^- \quad , & \quad x = z + \phi\,z^+ \\
    s^+ = \omega^+ + \frac{1}{2}\omega^-\,\phi^2 \quad , & \quad
    x^- = z^- + \phi\,z + \frac{1}{2}z^+\,\phi^2
  \end{aligned}
 \label{adapted1}
\end{equation}
Notice that \eqref{adapted1} is a well-defined coordinate system except at 
$\omega^-=0$, where it manifestly breaks down. When rewriting the embedding
equation \eqref{newembed} in terms of $\{z^\pm\,,\omega^\pm\,,z\,,\phi\}$, 
it becomes
\begin{equation}
  -2x^+x^- - 2\omega^+\omega^- + z^2 + \sum_{i=1}^{p-5}(x^i)^2 = -(\Ra)^2~,
 \label{eqb}
\end{equation}
so that it describes $\text{AdS}_{p-1}$ whenever $\omega^-\neq 0$. Since
$\xi=\partial_\phi$, $\phi$ becomes a compact dimension after the
identification. The metric of the corresponding quotient manifold 
\begin{equation}
  \begin{aligned}
    g_{\text{AdS}_p/\Gamma^{+-}} &= g_{\text{AdS}_{p-1}} 
    + [(\omega^-)^2 + (z^+)^2](d\phi)^2 + 2(z^+)^2
    d\phi \,d\left(\frac{z}{z^+}\right) \\
    &= g_{\tilde{\cM}} + \|\xi\|^2\left(d\phi + A_1\right)^2~.
  \end{aligned}
 \label{nullbrane}
\end{equation}
describes an $S^1$ fibration over some compact manifold $\tilde{\cM}$ with
non-trivial gauge field, whose field strength $F_2=d\,A_1$ is given by
\[
  F_2 = \frac{2}{[(\omega^-)^2 + (z^+)^2]^2}\left\{(\omega^-)^2\,
  dz^+\wedge dz - \omega^- \cdot (z^+)^2\, d\omega^-\wedge 
  d\left(\frac{z}{z^+}\right)\right\}~.
\]

It is easy to identify which is the nature of the conformal boundary in this
case. The analysis of symmetries given in the previous section teaches us that
the conformal boundary of $\text{AdS}_p/\Gamma^{+-}$ must be equal to the
conformal compactification of $\bR^{1,p-2}/\Gamma_{\text{null-brane}}$, where
$\Gamma_{\text{null-brane}}$ stands for the discrete action generated by
\[
  \xi_{\text{null-brane}} = R\partial_z + \xi^+~,
\]
acting non-trivially on $\bR^{1,3}$.
This statement can be checked as we did for null orbifolds in $\text{AdS}_p$.
For concreteness, we shall concentrate on the particular case $p=5$. Take the
local description \eqref{nullbrane} for $\text{AdS}_5/\Gamma^{+-}$ and
parametrise $\text{AdS}_4$ in global coordinates \eqref{global}. In particular,
\begin{equation*}
  \begin{aligned}
    z^+ = \frac{1}{\sqrt{2}}(u+x) &= \frac{\Ra}{\sqrt{2}}\cosh\rho
    \left(\cos\tau_{\text{AdS}} + \tanh\rho\,\Omega^x\right) \\
    \omega^- = \frac{1}{\sqrt{2}}(v-\omega) &= \frac{\Ra}{\sqrt{2}}\cosh\rho
    \left(\sin\tau_{\text{AdS}} - \tanh\rho\,\Omega^\omega\right) \\
    z &= \Ra\sinh\rho\,\Omega^z
  \end{aligned}
\end{equation*}
in such a way that $\Omega^i$ parametrise a unit 3-sphere. Extending our 
previous arguments, it can still be checked that there exists a conformal 
boundary at $\rho\to\infty$. The metric on such conformal boundary is
\begin{multline}
  \tilde{g} = -(d\tau_{\text{AdS}})^2 + g_{S^2} + \frac{1}{2}(d\phi)^2\left[
  (\sin\tau-\Omega^\omega)^2 + (\cos\tau_{\text{AdS}} + \Omega^x)^2\right] \\
  + 2\sqrt{2} d\phi \left[(\cos\tau_{\text{AdS}} + \Omega^x)\,d\Omega^z - 
  \Omega^z (d\Omega^x -\sin\tau_{\text{AdS}} d\tau_{\text{AdS}})\right] ~.
 \label{confbound1}
\end{multline}  
On the other hand, the metric on the null-brane vacuum 
$\bR^{1,3}/\Gamma_{\text{null-brane}}$
\begin{equation}
  g = -2dx^+dx^- + (dy)^2 + [R^2 + (x^+)^2](d\phi)^2 + 2d\phi\,
  \left(x^+\,dy -y\,dx^+\right)
\end{equation}
can be rewritten in terms of
\begin{equation*}
  \begin{aligned}
    x^{\pm} &= \frac{1}{\sqrt{2}}\left(\tau_f \pm r\,\Omega^x\right) \\
    y &= r\,\Omega^y ~.
  \end{aligned}
\end{equation*}
By standard procedures, its conformal compactification is given by
\begin{multline}
  g = -(d\tau_{\text{f}})^2 + g_{S^2} + \frac{1}{2}(d\phi)^2\left[
  2R^2(\cos\tau_{\text{f}}+\cos\theta)^2 +
  (\sin\tau_{\text{f}}+\Omega^x\,\sin\theta)^2\right]  \\
  + 2\sqrt{2}d\phi \left\{\sin\theta\,(\sin\tau_{\text{f}} +
  \Omega^x\sin\theta)\,d\Omega^y \right. \\
  \left. - \Omega^y\,(\sin\theta)^2 d\Omega^x +
  \Omega^y\,(\sin\tau_{\text{f}}\cos\theta\,d\theta - \cos\tau_{\text{f}}
  \sin\theta\,d\tau_{\text{f}})\right\}
 \label{confcomp1}
\end{multline}

One thus learns that using the identifications
\begin{equation}
  \begin{aligned}
    \tau_{\text{f}} &= \frac{\pi}{2} + \tau_{\text{ads}} \\
    \Omega^z &= \sin\theta\,\Omega^y \\
    \Omega^x &= \sin\theta\,\Omega^x \\
    \Omega^\omega &= \cos\theta 
  \end{aligned}
 \label{2dir}
\end{equation}
both, the conformal compactification \eqref{confcomp1} and the conformal 
boundary \eqref{confbound1} are equivalent. Notice that the angular
identifications in \eqref{2dir} match the two directions in which the null
rotations are defined in $\AdS_5$ with the two directions that the null-brane
discrete quotient defines on the conformal compactification of Minkowski
spacetime. 

When considering the above quotient in string/M-theory, the discussion of the
metric is as above. Concerning the field strengths in \eqref{fluxes}, one only
has to deal with the one in type IIB since the $\G^{+-}$ action can not be
defined in $\text{AdS}_4$ and it does not change $F_4$ for the 
$\text{AdS}_7\times\text{S}^4$ vacuum. Using the local adapted coordinate 
system for $\left(\text{AdS}_5/\G^{+-}\right)\times\text{S}^5$
\begin{equation}
  \begin{aligned}
    F_5 &= \frac{1}{2\Ra}\left\{\sqrt{(\omega^-)^2 + (z^+)^2} 
    \dvol\,\tilde{\cM}\wedge d\phi + \dvol\left(\text{S}^5\right)\right\} \\
    & =  \frac{1}{2\Ra}\left\{\|\xi\|\,
    \dvol\,\tilde{\cM}\wedge d\phi + \dvol\left(\text{S}^5\right)\right\}~.
  \end{aligned}
\end{equation}

Both $\left(\text{AdS}_5/\G^{+-}\right)\times\text{S}^5$ and
$\left(\text{AdS}_7/\G^{+-}\right)\times\text{S}^4$ preserve one quarter
of the spacetime supersymmetry. This is explicitly proved in the appendix, but
it is very easy to argue using the embedding $\text{AdS}_p\hookrightarrow
\bR^{2,p-1}$. We have constant Killing spinors $\varepsilon_0$ in 
$\bR^{2,p-1}$. The action of $\G^{+-}$ preserves those satisfying
\[
  \left(\G_{-x}+\G_{\hat{+}s}\right)\varepsilon_0 = 0~,
\]
where $\hat{+}$ stands for a lightlike direction in the vr-plane, whereas
$-$ lies in the uy-plane. Since both $\G_{-x}$ and $\G_{\hat{+}s}$ are
nilpotent and commute, the quotient preserves $\nu=1/4$ of the original
spacetime supersymmetry.

\section{The quest for non-perturbative string cosmology}

As explained in the introduction, the existence of a compact direction
along the worldvolume of the branes, opens up a very natural possibility.
Whenever its effective radius becomes smaller than the string scale, one is
entitled to apply a T-duality transformation, and move to the
T-dual description. Let us concentrate on D3-branes, from now on. We can read 
such a condition from the corresponding near horizon geometry (p=3) in
\eqref{cosmohor}
\begin{equation*}
  \frac{\text{u}^2}{\text{g}^2_{\text{YM}}\,\text{N}}\left(R^2 + (x^+)^2\right)
  \,<\,1~.
\end{equation*}

Before giving further details, let us summarise the different descriptions
that we have at our disposal in this particular scenario. We are analysing
the open string sector of string theory in a null-brane background. Extending
the formalism developed in \cite{LMS1} for the parabolic orbifold, and further
used for the null-brane in \cite{LMS2,fabinger}, one could in principle
study a perturbative analysis on the worldsheet of D3-branes in such vacuum.
The expected low energy effective field theory on the branes would be that
of d=1+3 SYM on the null-brane manifold. This is a gauge theory that can be
defined in terms of $\cN=4$ SYM by requiring the connection $A(X)$ and
the scalars transforming in the adjoint representation of $\SU(N)$ to
satisfy the boundary conditions
\begin{equation}
  \begin{aligned}
    \Phi^j (g_o\cdot X) &= \Omega (X)\Phi^j(X)\left[\Omega (X)\right]^{-1}
    \quad j=1,\dots ,6 \\
    A (g_o\cdot X) &= \Omega (X)\,A(X)\left[\Omega (X)\right]^{-1} - i\,
    d\Omega (X)\left[\Omega (X)\right]^{-1}
  \end{aligned}
 \label{projection}
\end{equation}
where $\Omega (X)$ stands for an $\SU(N)$ group element describing a 
gauge transformation, $X$ is a vector parametrising $\bR^{1,3}$ and $g_o$ is 
the generator of the null-brane discrete quotient
\begin{equation}
   X = \begin{pmatrix}
    z \\ x^+ \\ x  \\ x^- 
  \end{pmatrix} \quad , \quad
  g_o\cdot X =
  \begin{pmatrix}
    z + t\,R \\ x^+ \\ x + tx^+ \\ x^- + tx + \frac{1}{2}t^2 x^+
  \end{pmatrix} 
  ~.
\end{equation}
It would be interesting to check whether this gauge theory captures all the 
low energy effective dynamics of the open string sector at weak coupling. 

On the other hand, D-branes are massive charged objects which act as a source
for the various supergravity fields. Since these D-branes are excitations
over the null-brane vacuum, the corresponding classical D3-brane solution
is the one in \eqref{d3null} and mentioned in \cite{paper2}.
Relying on the intuition borrowed from the AdS/CFT correspondence, it seems
reasonable to establish some duality relation between the previously defined
gauge theory and type IIB in the background \eqref{cosmohor}, even though
the string theory dynamics in these backgrounds is still not understood.

It is at this stage that T-duality provides us with a bridge to go beyond
the Poincar\'e patch. Since the gravitational configuration we are dealing
with is locally isometric to the standard D3-brane solution, the forementioned
classical gravity description becomes reliable when the radius of curvature
of $\AdS_5$ and $\text{S}^5$ are large compared to the string scale
\begin{equation*}
  \frac{\Ra^4}{(\alpha^\prime)^2}\sim \text{g}_{\text{YM}}^2\,\text{N}
  \gg 1~.
\end{equation*}
Thus, one would expect T-duality to provide a reliable description whenever
\begin{equation}
  \text{u}^2\left(R^2 + (y^+)^2\right) < 
  (\text{g}_{\text{YM}}^2\,\text{N})^{1/2} \quad , \quad
   \text{g}_{\text{YM}}^2\,\text{N} \gg 1 ~.
\end{equation}
This is a time dependent condition, a fact which might not be that surprising
since energy in the dual gauge theory is not a conserved quantity. Even though
this condition restricts the regime of validity of the T-dual picture, what
remains true is that whenever it is reliable, it gives rise to a strongly
coupled type IIA configuration
\begin{equation*}
  e^\phi \sim \frac{g_s\,R}{\sqrt{\alpha^\prime}}\,\frac{
  (\text{g}_{\text{YM}}^2\,\text{N})^{1/4}}{\text{u}
  \left(R^2 + (y^+)^2\right)^{1/2}} \gg 1
\end{equation*}
which forces us to open up an extra eleven dimension, and look for the 
corresponding M-theory lift classical configuration. We shall provide such a 
lift before taking any near horizon limit, just because the corresponding
configuration is delocalised in two directions and it would not be that
illuminating. This is found to be
\begin{equation}
  \begin{aligned}
    g&= V^{1/3}\cdot L^{-2/3}\left(dx_\natural^2 + d\tilde{z}^2\right) +
    V^{1/3}\cdot L^{1/3}\left(dr^2 + r^2\,g_{S^5}\right) \\
    & + V^{-2/3}\cdot L^{1/3}\left\{2dy^+\,dy^- + dy^2 - L^{-1}\left(
    \frac{y}{R}dy^+ - \frac{y^+}{R}dy\right)^2\right\} \\
    C_3 &= V^{-1}\,dy^+\wedge dy^-\wedge dy + L^{-1}\left(  
    \frac{y^+}{R}dy -\frac{y}{R}dy^+\right)\wedge d\tilde{z}\wedge 
    dx_\natural~,
\end{aligned}
 \label{d=11}
\end{equation}
where we defined the scalar function 
\begin{equation*}
  L(y^+\,,R) = 1 + \left(\frac{y^+}{R}\right)^2 ~, 
\end{equation*}
$x_\natural$ stands for the eleventh dimensional coordinate and all quantities 
are measured in terms of the eleven dimensional Planck scale. 
The fact that supergravity does not take into account the physical effect
of any winding modes is, as usual, the responsible for the delocalisation
of \eqref{d=11}. As expected, not only the Lorentz group on the 
three dimensional M2-branes is broken due to the effect of the original
null-brane, but also the transverse Lorentz group is manifestly broken.
In particular, the two dimensional subspace spanned by $\{x_\natural\,,
\tilde{z}\}$ shrinks to zero size whenever $y^+\gg R$. In the other limit,
that is close to the ``neck'' of spacetime $(y^+\ll R)$, the metric in
\eqref{d=11} describes a three dimensional delocalised brane with a pp-wave 
propagating on it.

It is thus desirable to look for the most general eleven dimensional
supergravity configuration compatible with the isometries and supersymmetries
of our physical system. Its knowledge would teach us about the effect of the
original winding sector in the T-dual picture. This question is currently
under investigation. This problem is a very generic one and affects the T-dual 
descriptions of branes in both, fluxbrane and null-brane sectors in string
theory.

Such an eleven dimensional background would be expected to be dual to some sort
of large wave length approximation in the gauge theory. It would certainly be
very interesting to study the dynamics of such a gauge theory
\footnote{The following comments were suggested by Moshe Roszali.}. 
Due to its definition, correlation functions should be computable in terms of 
the known ones in $\cN=4$ SYM by the method of images. Two point functions with
times before and after the neck of spacetime and different spatial
wavelengths could teach us about the nature of the neck itself. For instance, 
it could be that for large wave lengths the penetration probability is
exponentially small. That would be interpreted as separation of the two 
branches of the cosmology.

Finally, we would also like to mention that the source of instability
discussed in \cite{fabinger,Horpol} may also exist in these scenarios,
due to the redshift of the energy. If so, it would be interesting
to understand its origin on the gauge theory side. Thinking of the black hole
as a thermal state, one could look for thermalization effects in the
two point function. We hope to come back to some of these issues in the
near future.

\medskip
\section*{Acknowledgments}
\noindent
The author would like to thank O. Aharony, M. Berkooz, N. Drukker,
JM. Figueroa-O'Farrill, N. Itzhaki, D. Kutasov, M. Roszali and A. Schwimmer 
for discussions on the topics presented in this work and for encouragement.
It is also a pleasure to thank the University of Edinburgh for hospitality 
during the intermediate stages of the present work. This research has been 
supported by a Marie Curie Fellowship of the European Community programme 
``Improving the Human Research Potential and the Socio-Economic knowledge 
Base'' under the contract number HPMF-CT-2000-00480.


\appendix

\section{Explicit supersymmetry analysis}

In this appendix, we determine the amount of supersymmetry preserved by
$\text{AdS}_p/\G^+$ and $\text{AdS}_p/\G^{+-}$. Since both, $\G^+$ and
$\G^{+-}$ just act on $\text{AdS}_p$, we shall decompose the ten or eleven
dimensional spinors into a convenient tensor product of spinors in AdS and
on the sphere. By default, we shall denote tne ones in AdS by $\varepsilon$.
These were determined in ~\cite{adsspinors}. Working in a coordinate system
where the AdS metric is described by
\begin{equation}
  g_{\text{AdS}} = (dr)^2 + e^{2\,\text{r}}\eta_{\alpha\beta}
  dx^\alpha\,dx^\beta~,
 \label{patch}
\end{equation}
it was proved that the Killing spinors $\varepsilon$ were satisfying
\begin{equation}
  \nabla_\mu\varepsilon = \frac{1}{2}\G_\mu\,\varepsilon \quad
  \mu=\alpha\,, \text{r}~,
\end{equation}
which is solved by 
\begin{equation}
  \varepsilon= e^{\text{r}/2}\,\varepsilon_+ \quad ; \quad
  \varepsilon= \left(e^{-\text{r}/2}+  e^{\text{r}/2} x^\alpha\G_{\ua}\right)
  \,\varepsilon_- \quad , \quad \G_{\underline{\text{r}}}\varepsilon_{\pm}
  = \pm\varepsilon_{\pm}~.
\end{equation}

The local condition of preservation of supersymmetry under the identification
$\G^+$ and $\G^{+-}$ requires the action of $\xi$ on the Killing spinors
$\varepsilon$ to vanish ~\cite{FOPflux}
\begin{equation}
  \cL_\xi\varepsilon = \nabla_\xi\varepsilon + \frac{1}{4}\nabla_a\xi_b
  \,\G^{ab}\varepsilon = 0 ~.
 \label{susycond}
\end{equation}  
Working out both terms, they can be written as :
\begin{equation}
  \begin{aligned}
    \nabla_\xi\varepsilon &= \frac{1}{2}e^{\text{r}}\left(\xi\cdot\G\right)
    \varepsilon \\
    \frac{1}{4}\nabla_a\xi_b \,\G^{ab}\varepsilon &= \left(\frac{1}{8}
    \left(\6_\alpha\5\xi_\beta-\6_\beta\5\xi_\alpha\right)\,\G^{\ua\ub} -
    \frac{1}{2}e^{\text{r}}\left(\xi\cdot\G\right)\,\G_{\underline{\text{r}}}
    \right)\varepsilon ~,
  \end{aligned}
\end{equation}
where $\xi\cdot\G = \xi^\alpha\,\G_{\ua}$, $\xi_\alpha = e^{2\text{r}}
\eta_{\alpha\beta}\xi^\beta= e^{2\text{r}}\5\xi_\alpha$ and $\{\G_{\umu}\,,
\G_{\unu}\}=2\eta_{\mu\nu}$.

It is manifest that for $\varepsilon=e^{\text{r}/2}\,\varepsilon_+$ condition
\eqref{susycond} reduces to the one in flat spacetime
\[
  \G_{-\underline{x}}\varepsilon = 0~.
\]
For the second kind of Killing spinors, the one involving $\varepsilon_-$,
splitting $\alpha=\{\ha\,,z\}$, $z$ being the direction in which $\xi$ acts
as a translation \footnote{The coordinates in \eqref{patch} describe AdS in
the Poincar\'e patch and it was proved in \eqref{match} that $\xi^-$ was mapped
to $R\6_z$ for some spacelike direction z transverse to the action generated
by $\xi^+$ .}
\begin{multline}
  \cL_\xi\varepsilon = \frac{1}{8}e^{-\text{r}/2}
  \left(\6_\alpha\5\xi_\beta-\6_\beta\5\xi_\alpha\right)\,\G^{\ua\ub}
  \varepsilon_- + e^{\text{r}/2}\,R\G_{\underline{z}}\,\varepsilon_-  \\
  +  e^{\text{r}/2}\left\{\left(\xi^{\ha}\cdot\G_{\hua}\right) +
  \frac{1}{8}\left(\6_{\ha}\5\xi_{\hb}-\6_{\hb}\5\xi_{\ha}\right)\,
  \G^{\hua\hub}\, x^{\hg}\cdot\G_{\hat{\underline{\gamma}}}\right\}
  \varepsilon_- = 0
 \label{middle}
\end{multline}
The first line is independent of $x^{\hg}$ whereas the second line is linear
in them. Thus, both must vanish independently. From the first one, we learn
\[
  \G_{-\underline{x}}\varepsilon_-=\G_{\underline{z}}\varepsilon_-=0 \quad 
  \Rightarrow \quad \varepsilon_-=0 \quad (R\neq 0)
\]
or
\begin{equation}
  \G_{-\underline{x}}\varepsilon_-=0 \quad (R=0) ~.
 \label{kl}
\end{equation}
When $R=0$, it can be shown using gamma matrices identities, the form of
$\xi^+$ and eq. \eqref{kl} that
\[
  \frac{1}{8}\left(\6_{\ha}\5\xi_{\hb}-\6_{\hb}\5\xi_{\ha}\right)\,
  \G^{\hua\hub}\, x^{\hg}\cdot\G_{\hat{\underline{\gamma}}}\varepsilon_- =
  - \xi^{\ha}\cdot\G_{\hua}\,\varepsilon_-
\]
so that the second line in \eqref{middle} is automatically satisfied
whenever the first is.

We sum up the conclusions of the above computation associating the generator
of the orbifold with the amount of supersymmetry $(\nu)$ that is
being preserved :
\begin{equation}
  \begin{aligned}
    \xi = \xi^+ \,\,(R=0) & \Rightarrow \,\,\nu=\frac{1}{2} \\
    \xi = \xi^++ \xi^- \,\,(R\neq0) & \Rightarrow \,\,\nu=\frac{1}{4} ~.
  \end{aligned}
\end{equation}



\begin{thebibliography}{99}

\bibitem{horowitz}
G.T. Horowitz and A.R. Steif, \emph{Singular string solutions with nonsingular
initial data}, Phys. Lett. \textbf{B258}, 91 (1991).

\bibitem{kt}
A.A. Tseytlin, \emph{Exact string solutions and duality}, 
  \texttt{arXiv:hep-th/9407099}; \par
C. Klimcik and A.A. Tseytlin, unpublished 
  (1994); A.A. Tseytlin, unpublished (2001). 

\bibitem{seiberg}
J. Khoury, B.A. Ovrut, N. Seiberg, P.J. Steinhardt and N. Turok, \emph{From Big
  Crunch to Big Bang},  Phys. Rev. \textbf{D65} (2002) 086007,
  \texttt{arXiv:hep-th/0108187}; \par
N. Seiberg, \emph{From Big Crunch to Big Bang : Is it possible ?},
  Talk given at the Francqui Colloquium 2001, \texttt{arXiv:hep-th/0201039}.  

\bibitem{paper1}
J.M. Figueroa-O'Farrill and J. Sim\'on, \emph{Generalised supersymmetric
  fluxbranes}, J. High Energy Phys. \textbf{12} (2001) 011,
   \texttt{arXiv:hep-th/0110170}.

\bibitem{vijay} 
V. Balasubramanian, S.F. Hassan, E. Keski-Vakkuri and A. Naqvi, 
  \emph{A Space-Time orbifold : A Toy Model for a Cosmological Singularity}, 
  \texttt{arXiv:hep-th/0202187}.

\bibitem{costa}
L. Cornalba and M. Costa, \emph{A new cosmological scenario in string theory},
  \texttt{arXiv:hep-th/0203031}.

\bibitem{nekrasov}
N. Nekrasov, \emph{Milne Universe, Tachyons and Quantum Group}, 
  \texttt{arXiv:hep-th/0203112}.

\bibitem{joan1}
J. Sim\'on, \emph{The geometry of null rotation identifications}, J. High
  Energy Phys. \textbf{06} (2002) 001, \texttt{arXiv:hep-th/0203201}.

\bibitem{LMS1}
H. Liu, G. Moore and N. Seiberg, \emph{Strings in a time-dependent orbifold},
  J. High Energy Phys. \textbf{06} (2002) 045, \texttt{arXiv:hep-th/0204168}.  

\bibitem{newcosta}
L. Cornalba, M. Costa and C. Kounnas, 
  \emph{A Resolution of the Cosmological Singularity with Orientifolds},
  Nucl. Phys. \textbf{B637} (2002) 378-394, \texttt{arXiv:hep-th/0204261}.

\bibitem{LMS2}
H. Liu, G. Moore and N. Seiberg, \emph{Strings in time-dependent orbifolds},
  \texttt{arXiv:hep-th/0206182}.

\bibitem{fabinger}
M. Fabinger and J. McGreevy, \emph{On Smooth Time-Dependent Orbifolds and
  Null Singularities}, \texttt{arXiv:hep-th/0206196}.

\bibitem{albion}
A. Lawrence, \emph{On the instability of 3d null singularities},
  \texttt{arXiv:hep-th/0205288}.

\bibitem{Horpol}
G.T. Horowitz and J. Polchinski, \emph{Instability of Spacelike and Null 
Orbifold Singularities}, \texttt{arXiv:hep-th/0206228}.

\bibitem{hebrew}
S. Elitzur, A. Giveon, D. Kutasov and E. Rabinovici, \emph{From big bang to big
  crunch and beyond}, J. High Energy Phys. \textbf{06} (2002) 017,
  \texttt{arXiv:hep-th/0204189}.

\bibitem{ben}
B. Craps, D. Kutasov and G. Rajesh, \emph{String Propagation in the Presence
  of Cosmological Singularities}, J. High Energy Phys. \textbf{06} (2002)
  053, \texttt{arXiv:hep-th/0205101}.

\bibitem{AFHSWick}
O. Aharony, M. Fabinger, G. Horowitz, and E. Silverstein, \emph{Clean
  time-dependent string backgrounds from bubble baths}, J. High Energy
  Phys. {\bf 07} (2002) 007, \texttt{arXiv:hep-th/0204158}.

\bibitem{BRinWick}
D. Birmingham and M. Rinaldi, \emph{Bubbles in Anti-de-Sitter Space},
 \texttt{arXiv:hep-th/0205246}. 

\bibitem{BRWick}
V. Balasubramanian and S. Ross, \emph{The dual of nothing},
  \texttt{arXiv:hep-th/0205290}.

\bibitem{CaiWick}
R. Cai, \emph{Constant curvature black hole and dual field theory},
  \texttt{arXiv:hep-th/0206223}.

\bibitem{GMWick}
A. Ghezelbash and R. Mann, \emph{Nutty bubbles}, \texttt{arXiv:hep-th/0207123}.

\bibitem{BLWWick}
A. Buchel, P. Langfelder, and J. Walcher, \emph{On time-dependent backgrounds 
  in supergravity and string theory}, \texttt{arXiv:hep-th/0207214}.

\bibitem{GSSbranes}
M. Gutperle and A. Strominger, \emph{Spacelike branes}, J. High Energy Phys.
  {\bf 04} (2002) 018, \texttt{arXiv:hep-th/0202210}.

\bibitem{CGGSbranes}
C. Chen, D. Gal'tsov, and M. Gutperle, \emph{S-brane solutions in supergravity
  theories}, Phys. Rev. {\bf D66} (2002) 024043,
  \texttt{arXiv:hep-th/0204071}.

\bibitem{KMPSbranes}
M. Kruczenski, R. Myers, and A. Peet, \emph{Supergravity {S}-branes},
  J. High Energy Phys. {\bf 05} (2002) 039, \texttt{arXiv:hep-th/0204144}.

\bibitem{DKSbranes}
N. Deger and A. Kaya, \emph{Intersecting {S}-brane solutions of {$D{=}11$}
  supergravity}, J. High Energy Phys. {\bf 07} (2002) 038,
  \texttt{arXiv:hep-th/0206057}.

\bibitem{SenRT}
A. Sen, \emph{Rolling tachyon}, J. High Energy Phys. {\bf 04} (2002) 048,
  \texttt{arXiv:hep-th/0203211}.

\bibitem{SenTE}
A. Sen, \emph{Time evolution in open string theory},
  \texttt{arXiv:hep-th/0207105}.

\bibitem{Senlast}
P. Mukhopadhyay and A. Sen, \emph{Decay of Unstable D-branes with Electric 
  Fields}, \texttt{arXiv:hep-th/0208142}.

\bibitem{adscft}
J.M. Maldacena, \emph{The Large N limit of superconformal field theories and
  supergravity}, Adv. Theor. Math. Phys. \textbf{2} (1998) 231,
  \texttt{arXiv:hep-th/9711200};\par
S.S. Gubser, I.R. Klebanov and A.M. Polyakov, \emph{Gauge theory correlators
  from noncritical string theory}, Phys. Lett. \textbf{B428} (1998) 105,
  \texttt{arXiv:hep-th/9802109};\par
E. Witten, \emph{Anti-de Sitter space and holography}, Adv. Theor. Path. Phys.
  \textbf{2} (1998) 253, \texttt{arXiv:hep-th/9802150}.

\bibitem{review}
O. Aharony, S.S. Gubser, J.M. Maldacena, H. Ooguri and Y. Oz, 
  \emph{Large N field theories, String theory and Gravity}, Phys. Rep.
  \textbf{323} (2000) 83--386, \texttt{arXiv:hep-th/9905111}.

\bibitem{paper2} 
  J.M. Figueroa-O'Farrill and J. Sim\'on, 
  \emph{Supersymmetric Kaluza-Klein reductions of M2 and M5-branes}, 
  \texttt{arXiv:hep-th/0208107}.

\bibitem{paper3} 
  J.M. Figueroa-O'Farrill and J. Sim\'on, 
  \emph{Supersymmetric Kaluza-Klein reductions of M-waves and MKK-monopoles}, 
  \texttt{arXiv:hep-th/0208108}.

\bibitem{hormarolf}
G.T. Horowitz and D. Marolf, \emph{A new approach to string cosmology},
  J. High Energy Phys. \textbf{07} (1998) 014,  \texttt{arXiv:hep-th/9805207}.

\bibitem{ads1}
K. Behrndt and D. L\"ust, \emph{Branes, Waves and AdS Orbifolds}, 
  J. High Energy Phys. \textbf{07} (1999) 019, \texttt{arXiv:hep-th/9805207}.

\bibitem{ads2}
B. Ghosh and S. Mukhi, \emph{Killing spinors and Supersymmetric AdS orbifolds},
  J. High Energy Phys. \textbf{10} (1999) 021,\texttt{arXiv:hep-th/9908192}.

\bibitem{cai1} 
R. Cai, \emph{Constant curvature black hole and dual field theory},
  \texttt{arXiv:hep-th/0206223}. 

\bibitem{BTZ} 
M. Ba\~nados, C. Teitelboim and J. Zanelli, Phys. Rev. Lett. 
  \textbf{69}, 1849 (1992).

\bibitem{henneaux}
M. Ba\~nados, M. Henneaux, C. Teitelboim and J. Zanelli, \emph{Geometry of the
  2+1 black hole}, Phys. Rev. \textbf{D48}, 1506 (1993),
  \texttt{arXiv: gr-qc/9302012}.

\bibitem{akisav}
A. Hashimoto and S. Sethi, \emph{Holography and String Dynamics in 
  Time-Dependent Backgrounds}, \texttt{arXiv:hep-th/0208126}.

\bibitem{jose1}
J.M. Figueroa-O'Farrill, \emph{On the supersymmetries on {A}nti-de~{S}itter 
  vacua}, Class. Quant. Grav. \textbf{16} (1999) 2043-2055, 
  \texttt{arXiv:hep-th/9902066}.

\bibitem{ppwave}
M. Blau, J.M. Figueroa-O'Farrill, C. Hull and G. Papadopoulos, \emph{Penrose
  limits and maximal supersymmetry}, Class. Quant. Grav. \textbf{19} (2002)
  L87-L95, \texttt{arXiv:hep-th/0201081}.

\bibitem{paper5}
J.M. Figueroa-O'Farrill and J. Sim\'on, \emph{Discrete quotients in AdS
  spaces}, to appear.

\bibitem{adsspinors}
H. L\"u, C.N. Pope and P.K. Townsend, \emph{Domain walls from anti-de Sitter
  spacetime}, Phys. Lett. \textbf{B391} (1997) 39, 
  \texttt{arXiv:hep-th/9607164}.

\bibitem{FOPflux}
J.M. Figueroa-O'Farrill and G. Papadopoulos, 
  \emph{Homogeneous fluxes, branes and a maximally supersymmetric solution 
  of {M}-theory}, J. High Energy Phys.
  \textbf{06} (2001), 036, \texttt{arXiv:hep-th/0105308}.


\end{thebibliography}
\end{document}